\documentclass[aps,10pt,preprint,showpacs,preprintnumbers,amsmath,amssymb,showkeys,superscriptaddress]{revtex4-1}

\usepackage{graphicx}
\usepackage{dcolumn}
\usepackage{bm}
\usepackage{lipsum}
\usepackage[most]{tcolorbox}
\usepackage[section]{placeins}
\usepackage{float}
\usepackage{mwe}

\begin{document}

\title{Large Rashba splittings in bulk and monolayer of BiAs}

\author{Muhammad Zubair} 
\email{mzubair@udel.edu}
\affiliation{Department of Physics and Astronomy, University of Delaware, Newark, DE 19716, USA }

\author{Igor Evangelista} 
\affiliation{Department of Materials Science and Engineering, University of Delaware, Newark, DE 19716, USA}

\author{Shoaib Khalid}
\affiliation{Princeton Plasma Physics Laboratory, P.O. Box 451, Princeton, New Jersey 08543, USA}

\author{Bharat Medasani}
\affiliation{Princeton Plasma Physics Laboratory, P.O. Box 451, Princeton, New Jersey 08543, USA}

\author{Anderson Janotti}
\email{janotti@udel.edu}
\affiliation{Department of Materials Science and Engineering, University of Delaware, Newark, DE 19716, USA}

\begin{abstract}
 Two-dimensional materials with Rashba split bands near the Fermi level are key to developing upcoming next-generation spintronics. They enable generating, detecting, and manipulating spin currents without an external magnetic field. Here, we propose BiAs as a novel layered semiconductor with large Rashba splitting in bulk and monolayer forms. Using first-principles calculations, we determined the lowest energy structure of BiAs and its basic electronic properties. Bulk BiAs has a layered crystal structure with two atoms in a rhombohedral primitive cell, similar to the parent Bi and As elemental phases. It is a semiconductor with a narrow and indirect band gap. The spin-orbit coupling leads to Rashba-Dresselhaus spin splitting and characteristic spin texture around the L-point in the Brillouin zone of the hexagonal conventional unit cell, with Rashba energy and Rashba coupling constant for valence (conduction) band of $E_R$= 137 meV (93 meV) and $\alpha_R$= 6.05 eV\AA~(4.6 eVÅ).  In monolayer form (i.e., composed of a BiAs bilayer), BiAs has a much larger and direct band gap at $\Gamma$, with a circular spin texture characteristic of a pure Rashba effect. The Rashba energy $E_R$= 18 meV and Rashba coupling constant $\alpha_R$= 1.67 eV{\AA} of monolayer BiAs are quite large compared to other known 2D materials, and these values are shown to increase under tensile biaxial strain.  
\end{abstract}


\maketitle
\section{Introduction} \label{sec:intro}

Spin-orbit coupling (SOC) caused by the Rashba effect is one of the most fundamental and important phenomena for the next generation of spintronic devices \cite{koralek2009emergence,walser2012direct,ohno1999spin, karimov2003high,muller2008spin,griesbeck2012strongly, ye2012growth,wang2013temperature} as it facilitates the generation, detection, and manipulation of spin current without external magnetic field \cite{soumyanarayanan2016emergent}.
The strength of the Rashba effect is quantified by three parameters, the Rashba energy ($E_R$), the Rashba momentum ($K_o$), and the Rashba coupling constant ($\alpha_R$) \cite{bychkov1984properties}, and these can be tuned by an external electric field \cite{soumyanarayanan2016emergent}. Spin-degenerate bands split into two parabolic bands, e.g. spin up and spin down, with the Rashba Hamiltonian given by $H_R$ = $\alpha_R$($\sigma$ x p).$\Hat{z}$. The band dispersion can be described by $E(k) = (\hbar^2k^2/2m) \pm  \alpha_R|k|$, where $m$ is the effective mass of the electron or hole.  This type of SOC is usually caused by lack of inversion symmetry due to a confining potential associated with an interface or surface \cite{bychkov1984properties,rashba1960properties,casella1960toroidal}, whereas the lack of inversion symmetry in the bulk is associated with the Dresselhaus type of spin-orbit coupling \cite{dresselhaus1955spin,d1986spin}. In many materials, the Rashba and Dresselhaus SOC effects are coupled, resulting in spin-splitting anisotropy that causes interesting phenomena such as the spin helix and long spin relaxation times observed in GaAs \cite{koralek2009emergence,walser2012direct,ohno1999spin,fang2017spin,karimov2003high,muller2008spin,griesbeck2012strongly,ye2012growth,wang2013temperature}.
\begin{figure}
\centering
\includegraphics[width=2.8in]{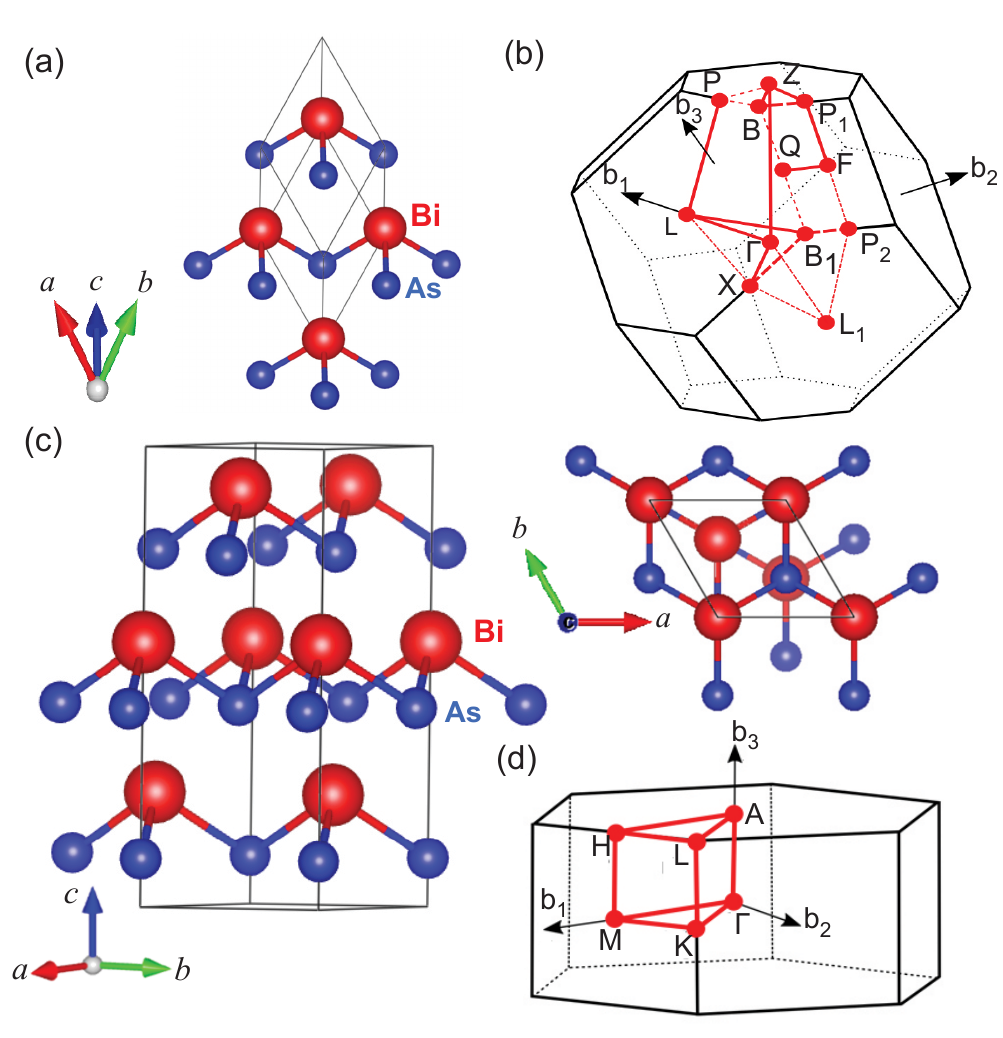}
\caption{Layered crystal structure of bulk BiAs, with R$\bar{3}$m space group, (a) represented the rhombohedral primitive cell containing two atoms and (b) the corresponding Brillouin zone, (c) the conventional hexagonal unit cell containing 6 atoms (side and top views), and (d) the corresponding Brillouin zone with the high symmetry points and directions. }
\label{fig1}
\end{figure}

Rashba spin-orbit splitting has been observed in spin-resolved and angle-resolved photoemission spectroscopy measurements of Au(111) \cite{lashell1996spin,hochstrasser2002spin,hoesch2004spin}, and in InAlAs/InGaAs \cite{nitta1997gate} and LaAlO$_3$/SrTiO$_3$\cite{caviglia2010tunable,zhong2013theory} interfaces, with modest values of the Rashba coupling constants $\alpha_R$ in the range 0.01-0.07 eV\AA~\cite{nitta1997gate,caviglia2010tunable,zhong2013theory}. 
Values of $E_R$ and $\alpha_R$ are larger in materials composed of heavy atoms \cite{manchon2015new}. For example,  $\alpha_R$= 0.55, 1.3, 0.33 and 3.05 eV\AA~were reported for the surface of heavy metals such as Bi\cite{koroteev2004strong}, Ir\cite{varykhalov2012ir}, Au\cite{lashell1996spin} and Bi/Ag(111) \cite{ast2007giant}, respectively. However, the metallic surface states in these materials make it complicated to use them in spintronics.

Two-dimensional (2D) semiconductors with strong Rashba effects would make great materials for devices such as spin field-effect transistors \cite{ahn20202d}.  A 2D Rashba semiconductor interfaced with an s-wave superconductor under broken time-reversal symmetry could also be used to build topological heterostructures to detect Majorana fermions \cite{sau2010generic,sau2010non,alicea2010majorana}, which can potentially be used to store and process quantum information in quantum computation \cite{sato2009topological,mourik2012signatures}. 
Among all 2D materials with strong SOC, those containing group-V elements such as bismuth (Bi), antimony (Sb), and arsenic (As) have received special attention in the investigation of Rashba effects \cite{takayama2014rashba,sheng2021rashba}. Bulk crystals of bismuth are semi-metallic with a small overlap of valence and conduction bands \cite{hofmann2006surfaces,hsu2019topology}.  A peculiar Rashba-type band splitting was observed on Bi(111) surface by high-resolution spin- and angle-resolved photoemission spectroscopy using Si(111) as substrate \cite{miyahara2012observation}. Sb and As are also semi-metals with the same crystal structure, but weaker SOC effects due to their lower atomic numbers.


\begin{figure}
\centering
\includegraphics[width=3.2in]{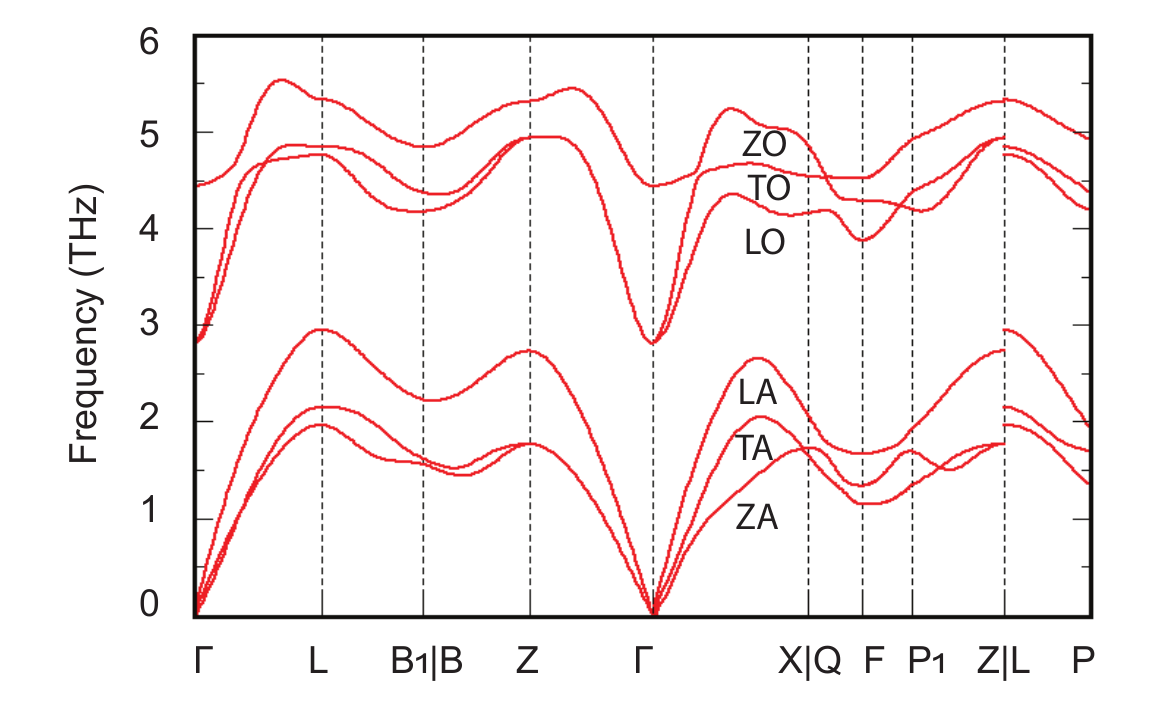}
\caption{Phonon dispersion along the high symmetry lines for the rhombohedral R$\bar{3}$m structure of the bulk phase of BiAs.}
\label{fig2}
\end{figure}
BiSb has also been recently investigated for the Rashba effect. Bulk BiSb is stable in the rhombohedral R$\bar{3}$m phase and also forms a layered structure, as in Fig.~\ref{fig1}. Density functional theory calculations reveal that BiSb is an indirect bandgap semiconductor with E\textsubscript{g}= 0.16 eV, with a band splitting at the L point in the hexagonal Brillouin zone, attributed to the SOC.
The spin-projected band structure in the H-L-H direction reveals the existence of Rashba-Dresselhaus-type spin splitting in this material \cite{singh2016investigation}.  The Rashba energy $E_R$ and the Rashba coupling constant $\alpha_R$ were calculated for both the conduction-band minimum (CBM) and the valence-band maximum (VBM), with values $E_R$=147 meV and $\alpha_R$= 10.4 eV\AA, and $E_R$=66 meV and $\alpha_R$=4.71 eV\AA, respectively \cite{singh2016investigation}.
The calculated Rashba parameters for the bulk BiSb is large compared with other bulk Rashba semiconductors such as BiAlO\textsubscript{3} ($E_R$ = 7-9 meV, $\alpha_R$= 0.4-0.7 eV\AA )\cite{da2016rashba}, BiTeI ($E_R$ = 100 meV, $\alpha_R$= 3.8 eV\AA) \cite{maass2016spin}, BiTeCl ($E_R$ = 19 meV, $\alpha_R$= 1.2 eVÅ )\cite{landolt2013bulk},  GeTe ($E_R$ = 227 meV, $\alpha_R$= 4.8 eV\AA )\cite{di2013electric}, SnTe ($E_R$ = 272 meV, $\alpha_R$= 6.8 eV\AA )\cite{lee2020emergence}, LiZnSb ($E_R$ = 21 meV, $\alpha_R$ = 1.82 eV\AA) \cite{narayan2015class} and KMgSb ($E_R$ = 10 meV, $\alpha_R$ = 0.83 eV\AA) \cite{narayan2015class}.

For the monolayer BiSb,  the calculated band structure shows a band gap of 0.37 eV (SOC included). 
The CBM, at $\Gamma$, displays a large spin splitting with $E_R$=13 meV and $\alpha_R$=2.3 eV\AA \cite{singh2017giant};  the VBM remains degenerate even with the inclusion of SOC. These Rashba parameters are very large compared to other 2D materials \cite{lashell1996spin,nitta1997gate,caviglia2010tunable,zhong2013theory,koroteev2004strong}.


In the present work, we use first-principles calculations to predict the structural and electronic properties of BiAs.
We find that BiAs also forms a layered structure with a rhombohedral R$\bar{3}$m space group, similar to BiSb. The results of the electronic structure calculations show that BiAs is a semiconductor with a small indirect band gap, with a SOC-induced splitting near the L point in the Brillouin zone of the hexagonal unit cell. For the bulk BiAs, we find $E_R$=137 meV and $\alpha_R$=6.05 eV\AA. As a monolayer, BiAs is a semiconductor with a direct band gap and Rashba splitting at $\Gamma$. The calculated spin texture confirms that monolayer BiAs shows a pure Rashba effect.
The presence of the giant Rashba spin splitting together with a large band gap makes this system of great interest for optoelectronic and spintronic applications. We demonstrated that the strength of the Rashba effect could be tuned by applying biaxial strain. Our results suggest that BiAs could be used for high-efficiency spin-field-effect transistors, optoelectronics, and spintronics devices.

\section{Computational Method} \label{sec:Com}

\begin{figure*}
\centering
\includegraphics[width=6in]{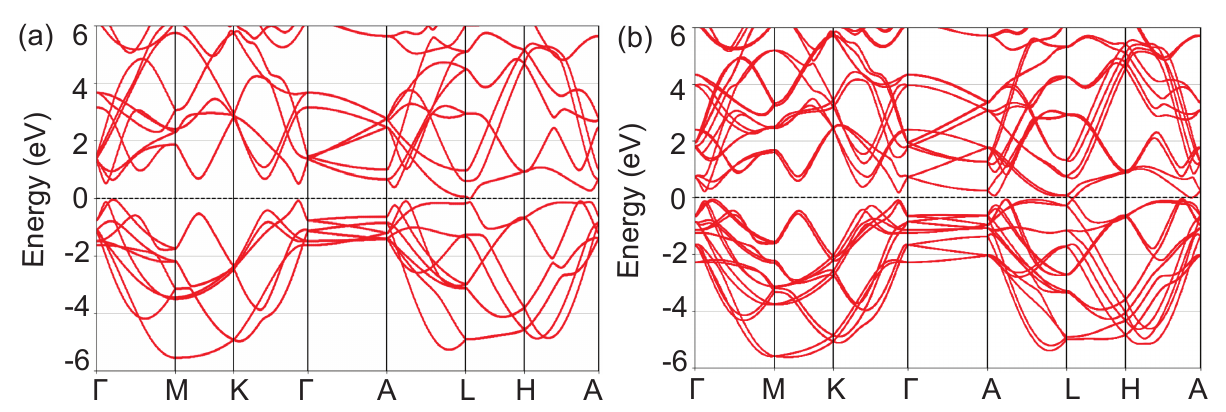}
\caption{Band structure of bulk BiAs (a) without spin-orbit coupling and (b) with spin-orbit coupling.}
\label{fig3}
\end{figure*}

Density functional theory (DFT) \cite{hohenberg1964inhomogeneous,kohn1965self} calculations as implemented in the VASP code \cite{kresse1996efficient} were employed to investigate the structural and electronic properties of BiAs. The interactions between the valence electrons and ions were treated using the projector augmented-wave (PAW) method \cite{kresse1999ultrasoft}, that include 5 valence electrons of Bi ($6s^26p^3$) and 5 valence electrons of As ($4s^24p^3$). We used the generalized gradient approximation(GGA) including van der Waals interactions \cite{PhysRevLett.77.3865} for determining equilibrium structures and the electronic structure. We also used the screened hybrid functional of Heyd-Scuseria-Ernzerhof (HSE06) \cite{heyd2003hybrid,heyd2004efficient} for determining band gaps based on the structures obtained using the GGA functional. Plane-wave basis set with a cutoff of 500 eV and an 8$\times$8$\times$8 k-mesh for the integrations over the Brillouin zone of the  2-atom rhombohedral primitive cell or equivalent were employed in all calculations. For convergence of the electronic self-consistent calculations, the total energy difference criterion was set to $10^{-6}$ eV, and atomic positions were relaxed until the Hellmann-Feynman residual forces were less than $10^{-4}$ eV/\AA. The effects of spin-orbit coupling (SOC) were included in all electronic band structure calculations.

To investigate the effect of in-plane biaxial strain for monolayer BiAs, we varied the $a$ and $b$ lattice vectors up to +6\% (expansion) while performing relaxation of all the atomic positions. A vacuum of thickness larger than 15 {\AA}  was employed along the $c$-axis to avoid any interaction between periodic images in the case of the monolayer BiAs. The PYPROCAR code was used to calculate the constant energy contour plots of the spin texture \cite{romeropyprocar}.

\begin{figure}
\centering
\includegraphics[width=2.2in]{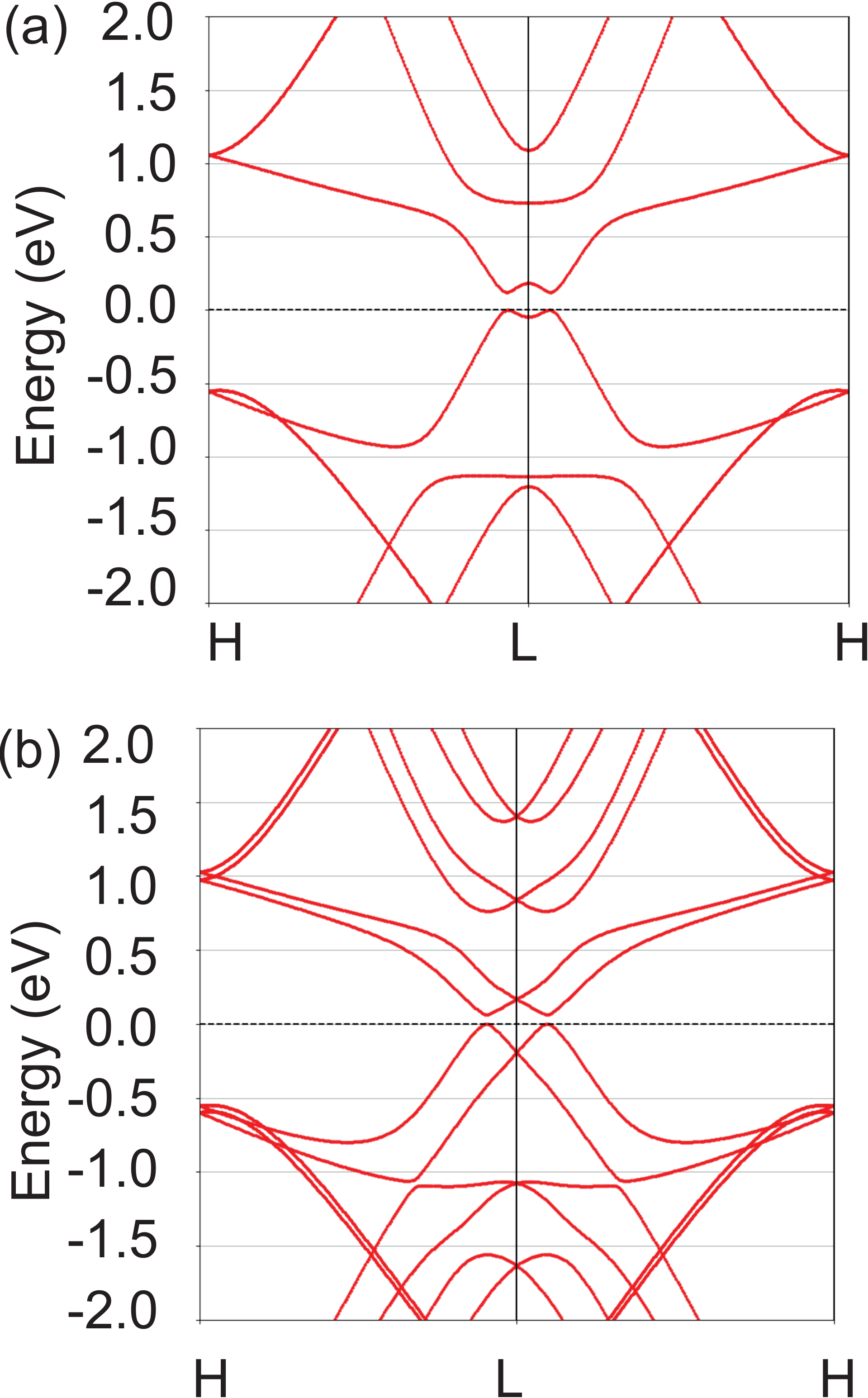}
\caption{Band structure of BiAs along the H-L-H direction of hexagonal Brillouin zone (a) without spin-orbit coupling and (b) with spin-orbit coupling.}
\label{fig4}
\end{figure}

\section{Results and Discussion}\label{sec:result}

We first carried out DFT calculations for BiAs in various possible crystal structures, including rocksalt, rhombohedral, hexagonal boron nitride, body-centered CsCl, and zinc blende. Only the rhombohedral R$\bar{3}$m crystal structure resulted in a negative formation enthalpy, of -80 meV eV/atom. In this structure, Bi and As atoms are stacked along the (111) direction of a rhombohedral primitive cell containing 2 atoms. It can also be described using a hexagonal unit cell with 6 atoms forming 3 BiAs bilayers that are weakly bonded by van der Waals interaction, as shown in Fig.~\ref{fig1}. The lattice parameters for the rhombohedral primitive cell are $a=b=c=4.283${\AA} and $\alpha$=$\beta$=$\gamma$=$58.2^{\circ}$. To check the dynamical stability of BiAs in the R$\bar{3}$m phase, phonon calculations using density functional perturbation theory (DFPT) were carried out using the PBE functional. The phonon band structure is shown in Fig.~\ref{fig2}. With two atoms in the unit cell, BiAs has six phonon branches, three lower frequency acoustic branches, and three higher frequency optical branches. From the phonon band structure, the longitudinal acoustic (LA)/longitudinal optical (LO) and transverse acoustic(TA)/transverse optical(TO), and out-of-plane acoustic (ZA)/out-of-plane optical (ZO) modes were identified. It can be clearly seen that there are no negative frequencies in the phonon band structure, attesting that BiAs is dynamically stable at the level of DFPT calculations.

\subsection{Electronic structure of bulk BiAs}

The conventional hexagonal unit cell (side and top views) along the Brillouin zone are shown in Fig.~\ref{fig1}. The lattice parameters for the hexagonal crystal structure are $a=b=4.167$~\AA, $c$=10.632{\AA} and the angles are $\alpha=\beta=90$$^{\circ}$,  $\gamma=120$$^{\circ}$.
The band structure of conventional hexagonal unit cell of BiAs without spin-orbit coupling (SOC) and with spin-orbit coupling are shown in Fig.~\ref{fig3} using the PBE functional. The band gap occurs along the H-L direction, closer to L.
The band gap without spin-orbit interaction is 45 meV, and with spin-orbit coupling is 42 meV.As the PBE functional underestimates the band gap, we employ Hyed-Scuseria-Ernzerhof (HSE06) hybrid functional to predict the band gap, which is most often comparable with the experimental results for many materials. \cite{sham1983density,godby1986accurate,sham1985density}
The band gap without and with SOC using HSE06 are 115 meV and 67 meV, respectively. The electronic band structure showed the presence of spin splitting around L point of the hexagonal Brillouin zone. Bahramy {\em et al.} proposed three conditions for the existence of a giant Rashba effect in bulk: (i) large spin-orbit coupling (SOC) in an inversion-asymmetric system, (ii) a narrow band gap, and (iii) the presence of VBM and CBM of symmetrically same character \cite{bahramy2011origin}. The band structure plotted along the H-L-H direction of the hexagonal Brillouin zone shows that the bulk BiAs crystal satisfies all these three conditions and consequently exhibits a very large Rashba effect, as seen in Fig.~\ref{fig4}(b). The presence of Rashba spin splitting is attributed to the strong spin-orbit coupling of the Bi atoms.
\begin{figure*}
\centering
\includegraphics[width=6.5in]{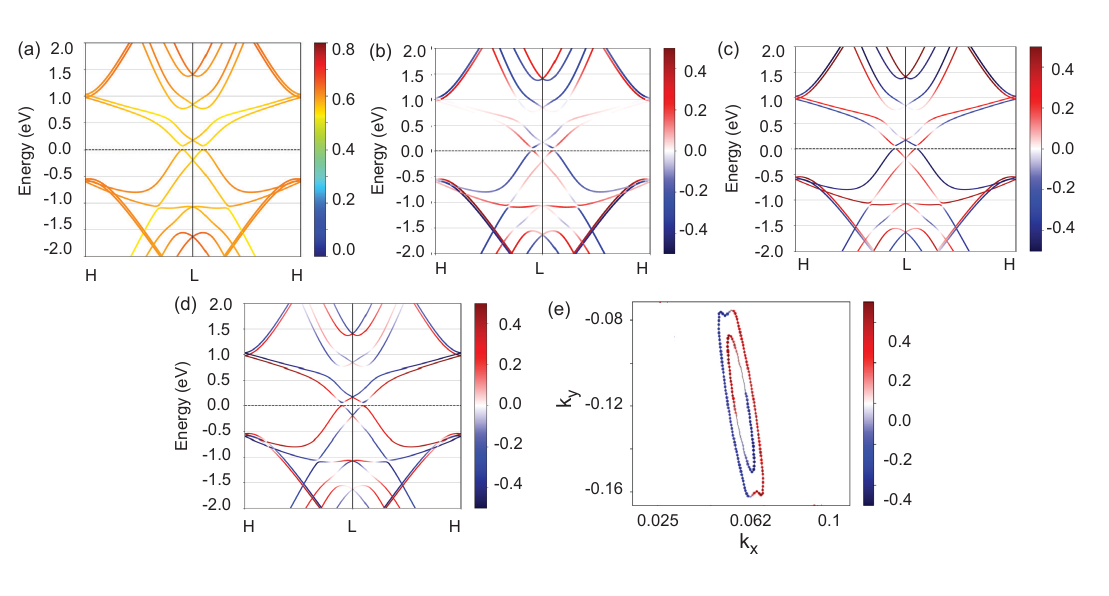}
\caption{ Spin-resolved band structure (S) of BiAs in H-L-H direction, (b) Spin projected band structure (S\textsubscript{z}) of BiAs in H-L-H direction, (c) Spin-projected band structure (S\textsubscript{y}) of BiAs in H-L-H direction, (d) Spin projected band structure (S\textsubscript{z}) in H-L-H direction, (e) Spin texture around L point of the Brillouin zone.}
\label{fig5} 
\end{figure*}

\begin{figure}
\centering
\includegraphics[width=2.9  in]{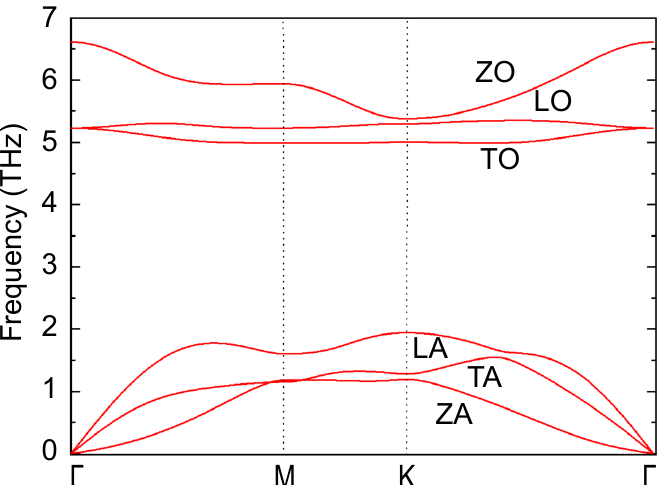}
\caption{Calculated phonon band structure of monolayer BiAs along high symmetry points.}
\label{fig6}
\end{figure}

\begin{figure*}
\centering
\includegraphics[width=5in]{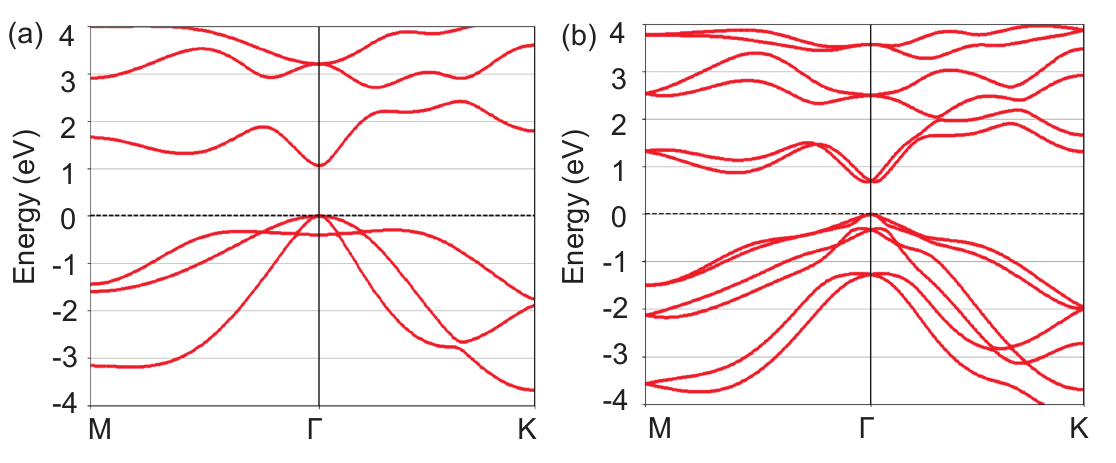}
\caption{ Band structure of monolayer BiAs (a) without spin-orbit coupling and (b) with spin-orbit coupling.}
\label{fig7}
\end{figure*}

\begin{figure*}
\centering   
\includegraphics[width=7in]{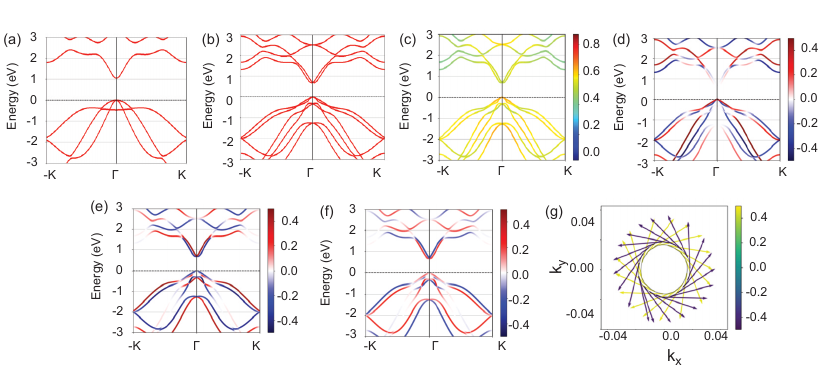}
\caption{(a) Band structure of monolayer BiAs without spin-orbit coupling in -K-$\Gamma$-K direction of Brillouin zone, (b)Band structure of monolayer BiAs with spin-orbit coupling in -K-$\Gamma$-K direction of Brillouin zone, (c) Spin band structure of monolayer BiAs in -K-$\Gamma$-K direction of Brillouin zone, (d) Spin projected band structure (S\textsubscript{z}) of monolayer BiAs in -K-$\Gamma$-K direction of Brillouin zone, (e), Spin projected band structure (S\textsubscript{x}) of monolayer BiAs in -K-$\Gamma$-K  of Brillouin zone, (f) Spin projected band structure (S\textsubscript{y}) in -K-$\Gamma$-K direction of Brillouin zone of monolayer BiAs, (g) Spin texture around $\Gamma$-point of the Brillouin zone of monolayer BiAs.}
\label{fig8}
\end{figure*}

The spin-resolved band structures showing the total spin (S) and spin projected on the cartesian coordinates S\textsubscript{x}, S\textsubscript{y} and S\textsubscript{z} along the H-L-H direction are shown in Fig.~\ref{fig5}. The results show significant spin-density located at VBM and CBM. The spin splitting of bands are observed in all three components of spin, indicating the presence of both Rashba- and Dresselhaus-type SOC.  The calculated Rashba energy $E_R$=93 meV and the Rashba coupling constant $\alpha_R$=4.6 eV{\AA} for the conduction band, and $E_R$=137 meV and $\alpha_R$=6.08 eV{\AA} for the valence band are large compared to other bismuth-containing materials such as BiTeI, BiTeBr, BiTeCl, and BiAlO\textsubscript{3}, which fall in the ranges 7.3-100 meV and 0.39-3.8 eV\AA. \cite{ishizaka2011giant,da2016rashba,xiang2015observation,martin2017experimental}
The shape of the spin texture in k\textsubscript{x}-k\textsubscript{y} plane and centered on the L is elliptical rather than circular as expected for a pure Rashba effect, as shown in Fig.~\ref{fig5}(e). This indicates the contribution of the Dresselhaus SOC effect, consistent with the lack of inversion symmetry in the BiAs crystal structure.

\subsection{Electronic structure of monolayer BiAs}

The Bi atoms form covalent bonds with the As in the monolayer BiAs, as shown in Fig.~\ref{fig1}(b), forming a honeycomb lattice geometry with three-fold rotational symmetry in the P3m1 space group.  The calculated equilibrium lattice parameters are $a=b=3.982$~{\AA} and $\gamma=120$$^{\circ}$.  The Bi-As bond length is 2.76~\AA, buckling height is 1.532~\AA, and the As-Bi-As angle is 92.2$^{\circ}$.  

\begin{figure}
\centering
\includegraphics[width=3.2in]{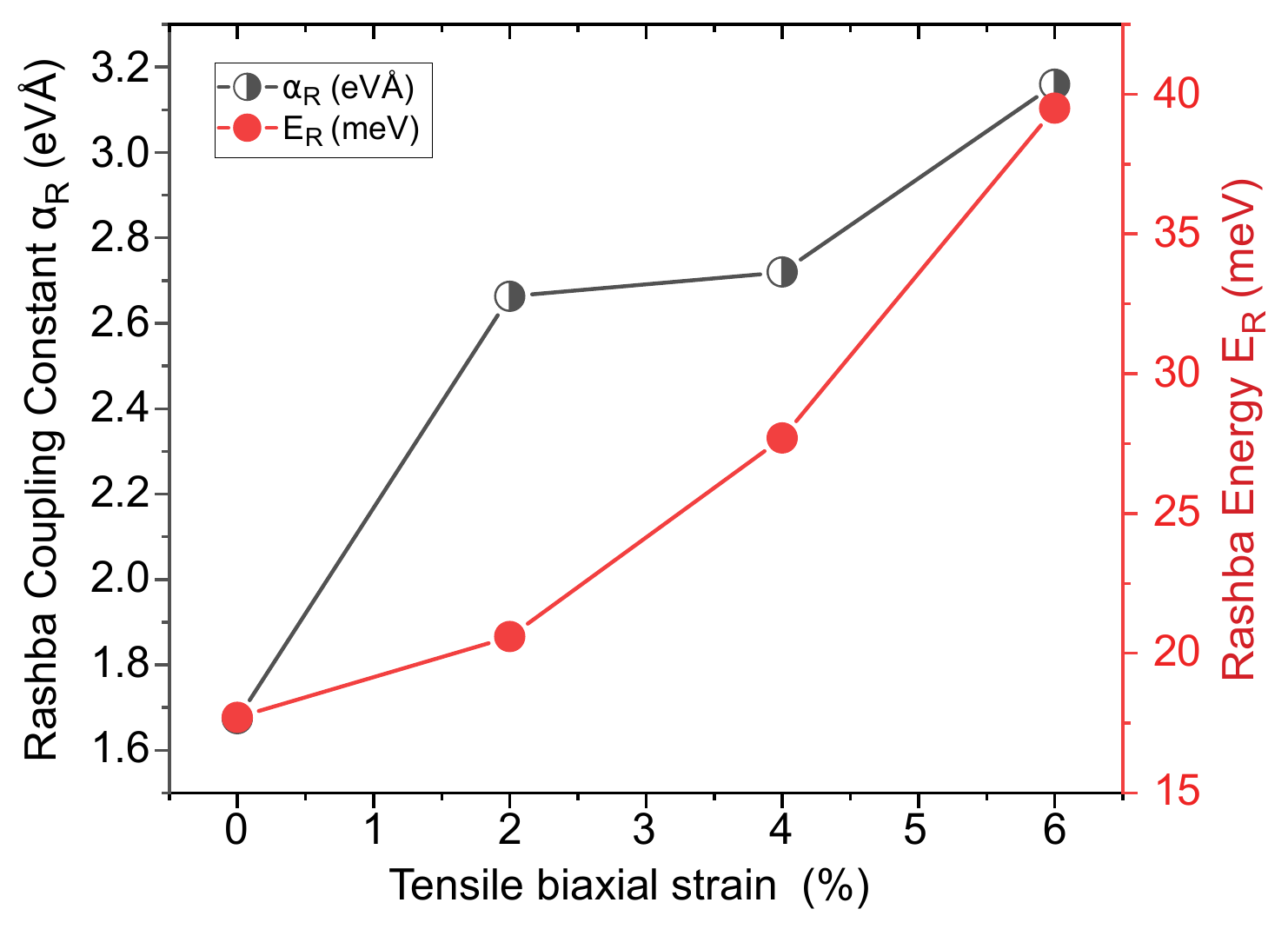}
\caption{Variation in Rashba parameters with the tensile biaxial strain in monolayer BiAs.}
\label{fig9}
\end{figure}

The phonon dispersion along the high-symmetry lines in the hexagonal Brillouin zone is shown in Fig.~\ref{fig6}. The longitudinal acoustic (LA) and transverse acoustic (TA) represent the in-plane vibrations. The flexural acoustic (ZA) band represents out-of-plane vibrations. It can be clearly seen that there are no negative frequencies in the phonon band structure, attesting that BiAs monolayer is dynamically stable at the level of DFPT calculations. The optical and acoustic modes are well separated from each other by more than 102 cm$^{-1}$, indicating a good optical response for monolayer BiAs.

The electronic band structure of the monolayer BiAs is shown in Fig.~\ref{fig7}. The band structure without SOC reveals a semiconducting nature with the presence of a direct band gap of 1.06 eV at the $\Gamma$ point. The band structure of spin-orbit coupling shows the direct band gap of 0.7 eV at $\Gamma$. This is in contrast to the much smaller and indirect band gap of bulk BiAs. 

The conduction band near the CBM at $\Gamma$ displays a significantly larger dispersion than the valence band, suggesting that electrons are bound to have significantly higher mobility than holes. As the PBE functional underestimates the band gap, we employ Hyed-Scuseria-Ernzerhof (HSE06) hybrid functional to predict the band gap. 
The band gap using HSE06 with SOC is 1.30 eV (1.72 without SOC). 

The band structure of monolayer BiAs along $-$K-$\Gamma$-K is shown in Fig.~\ref{fig8}(a). The spin splitting of the conduction band around $\Gamma$ is an indication of Rashba-type SOC. This is verified by inspecting the spin-resolved band structures in Fig.~\ref{fig8}(c)-(f).
The results show significant spin density located both at the valence band as well as the conduction band. For the conduction band, by comparing the projection on S\textsubscript{x}, S\textsubscript{y} and S\textsubscript{z}, we can clearly see the contributions from of S\textsubscript{x}  and S\textsubscript{y} but not S\textsubscript{z}, which is a clear indication of a pure Rashba effect. 
For the conduction band, we find $E_R$=18 meV and $\alpha$\textsubscript{R}=1.7 eV\AA. The spin texture around $\Gamma$ shown in Fig.~\ref{fig8}(g) has a circular shape, which is also an indication of the pure Rashba-type spin splitting present in monolayer BiAs.

\subsection{Effect of biaxial strain on the electronic structure of monolayer BiAs} 

One of the main characteristics of 2D materials is that they support large strains and can be easily stretched by placing them on textured substrates or flexible substrates that can be easily bent, leading to regions of uniaxial or biaxial strain.  These strained regions possibly favor charge localization and can attract or repel defects. As a demonstration of how strain impacts the electronic structure of BiAs, we studied monolayer BiAs under biaxial tensile strain up to +6\%. Our results, displayed in Fig.~\ref{fig9}, show that the Rashba energy $E_R$ and the Rashba coupling constant $\alpha_R$ increase with tensile strain up to 6\%, yet the material remains semiconducting. More specifically, $E_R$ and $\alpha_R$ increases from 17.7 meV and 1.67 eV\AA~ to 39.5 meV and 3.16 eV\AA, respectively, by applying a biaxial tensile strain of 6\%. 
These findings highlight the significant tunability of Rashba parameters by applying biaxial strain on the monolayer BiAs.

\subsection{Conclusions}

Using first-principles calculations, we identified BiAs as a 2D layered semiconductor with large Rashba splitting in the bulk and monolayer forms. The bulk BiAs is an indirect narrow band gap semiconductor stable in the rhombohedral crystal structure like the parent Bi and As compounds. The spin-orbit coupling results in Rashba-Dresselhaus type spin splitting around the L point of the Brillouin zone of the conventional hexagonal unit cell of the bulk form. The monolayer BiAs is a quasi-direct band gap semiconductor with a circular spin texture characterizing a pure Rashba effect. The Rashba parameters can be significantly enhanced by applying tensile biaxial strain in the monolayer. The existence of large Rashba effects in bulk and monolayer make BiAs a promising material for spin-field-effect transistors and optoelectronic devices in spintronics.

\section*{Acknowledgements}
\vspace{-1mm}
This work was supported by the National Science Foundation award \#OIA-2217786, and the use of Bridges-2 at PSC through allocation DMR150099 from the Advanced Cyberinfrastructure Coordination Ecosystem: Services \& Support (ACCESS) program, which is supported by the National Science Foundation grant nos.~2138259, 2138286, 2138307, 2137603, and 2138296, and the DARWIN computing system at the University of Delaware, which is supported by the NSF grant no.~1919839. S.K. acknowledges funding from the LDRD Program (Grant No.
PPPL-132) at Princeton Plasma Physics Laboratory under U.S. Department of Energy Prime Contract No. DE-
AC02-09CH11466.


\bibliography{Bib_ReV}
\end{document}